\documentstyle[prl,twocolumn,aps,epsf]{revtex}

\begin{document}
\draft
\twocolumn[\hsize\textwidth\columnwidth\hsize\csname@twocolumnfalse%
\endcsname

\title{Localized to extended states transition for two interacting \\
particles in a two-dimensional random potential}

\author{M. Ortu\~no and E. Cuevas}

\address{Departamento de F{\'\i}sica, Universidad de Murcia,
E-30071 Murcia, Spain.}

\date{\today}

\maketitle

\begin{abstract}
We show by a numerical procedure that a short-range interaction
$u$ induces extended two-particle states in a two-dimensional
random potential.
Our procedure treats the interaction as a perturbation
and solve Dyson's equation exactly in the subspace of doubly 
occupied sites. 
We consider long bars of several widths and extract the macroscopic
localization and correlation lengths by an scaling analysis of
the renormalized decay length of the bars. For $u=1$, the 
critical disorder found is $W_{\rm c}=9.3\pm 0.2$, and the critical 
exponent $\nu=2.4\pm 0.5$. For two non-interacting particles we do not
find any transition and the localization length is roughly
half the one-particle value, as expected.
\end{abstract}

]
\narrowtext

The interplay of disorder and interactions in electronic systems has
been studied intensively within the last two decades \cite{Fi83,LR85}.
Recent experimental results by Kravchenko {\it et al.} \cite{KS96}
have presented strong evidence for a metal-insulator transition in 
two-dimensional (2D) high mobility Si
metal-oxide-semiconductor field-effect transistors, and have generated 
a great deal of interest in the problem of the existence of a metallic
state in 2D.
These results have been confirmed by other workers employing different
materials and designs \cite{PF97}. All these experiments show clear
indications that strong electron-electron interactions partially suppress 
the quantum interference effects responsible for localization.
At the same time, the scaling theory of localization including the 
combined effects of disorder and interactions predicts that a 2D system
may remain metallic even in the limit of zero temperature \cite{Fi83}.

Direct numerical simulations of the problem are extremely difficult 
and, at present, we have to conform with solving the simplest related
problem, that of just two interacting particles (TIP) in a random 
potential.

In one-dimensional systems, this TIP problem has attracted a lot of 
attention since the original works of Dorokhov \cite{Do90} and 
Shepelyansky \cite{Sh94a}.
The problem has been approached from different points of view:
by using a Thouless type block-scaling picture \cite{Im95},
by mapping the TIP problem onto a random matrix problem \cite{JS95},
by direct numerical approaches based on the time evolution of wave
packets \cite{Sh94a,Sh94}, by transfer matrix methods \cite{FM95},
by Green function approaches \cite{OW96,SK97}, by exact
diagonalization \cite{WM95} and by a decimation method \cite{LR98}.
The previous works coincide on
the existence of a coherent pair propagation enhancement. 

In this Letter we perform a numerical calculation of the TIP problem
in a 2D random potential. Our main aim is to establish the existence of
extended coherent two-particle pairs.
Although in principle this is against the accepted one-particle scaling
picture, it should not be very surprising, since 2D is the critical 
dimension for localization. The consideration, for example, of 
spin-orbit coupling effects resulted in a metal-insulator
transition\cite{Ev95}.

Our algorithm combines an exact implementation of von Oppen
{\it  et al.} approach\cite{OW96} with the scaling procedure of MacKinnon 
and Kramer\cite{MK83} for the study of critical properties of disordered
systems. We consider long samples and calculate their decay lengths as 
a function of width and disorder energy. 
These data support the assumptions made in the scaling theory
and prove the existence of a transition from localized to extended
states at a finite critical disorder. In adition, to strengthen the 
validity of our calculations, we performed the
same analysis for two non-interacting particles and obtained the
expected localization length, i.e., half the value of the one-particle
localization length.

Due to the computational effort involved we concentrate in the case of
bosons with an on-site interaction, although we expect that our main
conclusion is equally valid for fermions. A small test of the dependence
of the renormalized decay length with the width of the sample
clearly shows the same trend for the existence of a 
transition as for bosons.

We consider a system of length $L$ and width $M$ described by the
standard Anderson-Hubbard hamiltonian for two spinless particles
\begin{eqnarray}
H & = & t\sum_{\{i,k\},j} |i,j\rangle \langle k,j| 
+t\sum_{i,\{j,l\}} |i,j\rangle \langle i,l| \nonumber \\
& & + \sum_{i,j} |i,j\rangle (\epsilon_i+\epsilon_j)\langle i,j|+U 
\equiv H_0+U \;,
\end{eqnarray}
where $i$ (and $j$, $k$, $l$) labels the $L \times M$ sites of a 
square lattice, and $\epsilon_i$ is the random site energy chosen 
from a box distribution with interval $[-W/2, W/2]$. 
$\{i,k\}$ (and $\{j,l\}$) indicates that the index $k$ ($l$) runs over 
the nearest neighbor sites of $i$ ($j$). The hopping matrix
element $t$ is taken equal to $-1$ and the lattice constant equal to
1, which sets the energy and length scales, respectively. 
We choose an on-site interaction with matrix elements
$\langle i,j|U|k,l\rangle=u \delta_{i,k}\delta_{j,l}\delta_{i,j}$,
and use lateral periodic boundary conditions.

To obtain the two-particle decay length $\xi$ of the
hamiltonian\ (1) we focus on the two-particle GF
\begin{equation}
G = (E - H_0 - U)^{-1}\;.
\end{equation}
The full GF satisfies Dyson's equation
\begin{equation}
G = G_0 + G_0UG\;,\label{dyson}
\end{equation}
where $G_0$ is the two-particle GF in the absence of interactions.
The eigenvectors and eigenvalues of the one-particle problem are enough 
to construct $G_0$ \cite{OW96}. 
von Oppen {\it et al.} \cite{OW96} noted that for a local interaction we 
can obtain $G$ very efficiently by projecting onto the subspace of 
doubly occupied sites. 
This is equivalent to solving first the non-interacting
case and considering the interaction as a perturbation, only acting on 
the subspace of doubly occupied sites.
We will refer with a tilde to the matrices restricted
to this subspace.  
Solving Eq.\ (\ref{dyson}) for $\widetilde{G}$, and taking into
account that $\widetilde{U}=u\openone$, we obtain
\begin{equation}
\widetilde{G}=(\openone-u \widetilde{G}_0)^{-1} \widetilde{G}_0\;.
\end{equation}
This expression can be evaluated exactly by inverting
matrices of range equal to the system size, $L \times M$.
These matrices are full, all elements are relevant, and their inversion 
cannot be alleviated by matrix transfer methods.

\begin{figure}
\begin{picture}(220,200) (-65,-50)
\epsfbox{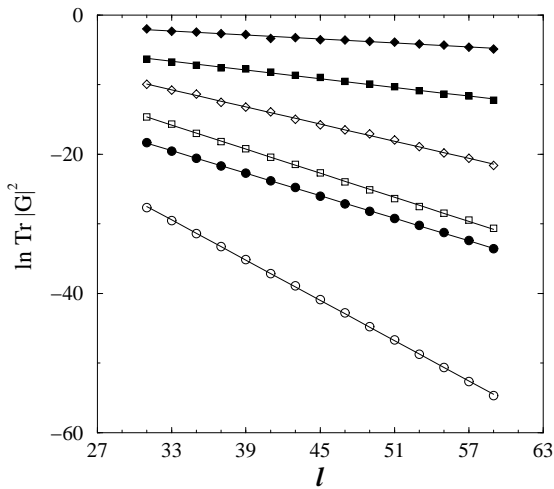}
\end{picture}
\caption{$\ln {\rm Tr}\, |\widetilde{G}|^2$ as a function of $l$ for
$W=6$, and M=2 (circles), 4 (squares) and 6 (diamonds).
Solid symbols correspond to $u=1$ and empty symbols to $u=0$.
The straight lines are fits from which we calculate the corresponding
decay lengths.
\label{fig.1}}
\end{figure}

Let us call $\widetilde{G}(m_1,n_1;m_2,n_2)$ to the matrix element of 
the GF between an initial (doubly occupied) site of coordinates 
$(m_1,n_1)$, and a final (doubly occupied) site of coordinates $(m_2,n_2)$. 
For a given strip of size $L \times M$ we calculate the following trace
\begin{equation}
\ln{\rm Tr}\,|\widetilde{G}(l)|^2\equiv \langle \ln\sum_{i,j} 
|\widetilde{G}(1,i;l,j)|^2 \rangle \;,\label{tra}
\end{equation}
with $l\le L$, and where $\langle\, \rangle$ denotes an average over the 
disorder realizations. We ensure that $L$ is large 
enough to get a linear exponential decay of the trace as a function of $l$, 
for any disorder $W$ and width $M$ considered. 
Once we reach the exponential 
regime, we fit the data in this regime to a straight line, whose slope 
$\alpha$ is related to the two-particle decay length $\xi_M$ through 
$\xi_M=-2/\alpha$.

In Fig.\ 1 we show $\ln {\rm Tr}\, |\widetilde{G}|^2$ as a function of 
the length $l$ for $M=2$ (circles), 4 (squares) and 6 (diamonds).
Solid symbols correspond to the interacting case and empty symbols to 
$u=0$. Each data is obtained by  averaging over at least 300 
configurations. The disorder strength is $W=6$, which for the 
interacting case already lies in the extended regime, as we will see. 
We can note that a linear exponential decay is well established in all
cases considered. Strictly speaking our results constitute a lower
bound for the decay length, but the quality of the exponential decay 
implies that a larger decay length would have a very small weight and so
would correspond to very rare two-particle states. To further check the
validity of our results, we will apply the same procedure to 
non-interacting particles and we will see that we obtain the expected
results as compared with transfer matrix calculations with very long
bars \cite{MK83,MK81}, which are not feasible for the interacting problem.

Finite-size scaling analysis\cite{MK83,MK81} states that the
renormalized decay length $\xi_M/M$ is a function of a single 
parameter $\xi/M$,
\begin{equation}
\xi_M/M=f(M/\xi)\;.\label{sca}
\end{equation}
The scaling parameter $\xi$ is the two-particle localization length in
the localized regime, and the two-particle correlation length in the
extended regime.
Equation (\ref{sca}) implies that in a log-log plot of $\xi_M/M$ versus
$M$ all data should collapse in a common curve when translated by
an amount $\ln \xi(W)$ along the horizontal axis.
This curve has a single branch when there is no transition, while it
develops two separate branches when a transition is present.
The main aim of this paper is to discern whether $\xi_M/M$ collapses 
into a single or into a double branched curve.

In Fig.\ 2, we show  the raw data for 
$\xi_M /M$ as a function of the system width $M$ on a double
logarithmic scale for different values of the disorder. The on-site 
energy is $u=1$ and the disorder energies range between $W=6$ and 15,
as indicated in the figure.
All data were obtained by averaging over a number of disorder
realizations ranging between 300, for the largest $M$, and 1000, for
$M=2$. We consider the center of the band ($E=0$), and a length
$L=62$. The fact that $\xi_M /M$ increases with $M$ for small values
of $W$, while it decreases for large values of $W$ is a clear sign of
the presence of a transition.

A scaling analysis of the data shown in Fig.\ 2 is depicted in
Fig.\ 3, where we have overlapped all points on one curve within the 
accuracy of the data, by shifting
the data horizontally by a disorder dependent amount, which is determined
by a least-square fit procedure \cite{MK83}.
Fitting the data set for $W=15$ and $M$ between 4 and 10 to the
form $\xi_M=\xi +A/M$ we obtain the localization length
for this disorder $\xi(15) =2.1 \pm 0.1$, which enables us to establish
the absolute scale of $\xi (W)$. 
As there is practically no overlap between the two branches, we have to
obtain the absolute scale for the upper branch by assuming that the
scaling parameter diverges symmetrically from above and bellow at the 
transition. The existence of two branches is a
clear indication of a transition. We will see later on
how for the non-interacting case we obtain only one branch, as expected.

\begin{figure}
\begin{picture}(220,200) (-55,-50)
\epsfbox{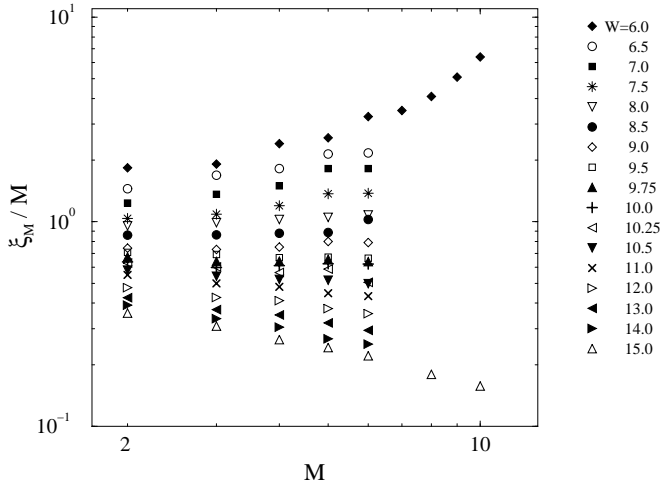}
\end{picture}
\caption{$\xi_M/M$ versus $M$ on a double
logarithmic scale for two-interacting particles with $u=1$ and for the 
indicated values of the disorder parameter $W$.
\label{fig.2}}
\end{figure}

In order to obtain the critical disorder $W_{\rm c}$ and the critical
exponent $\nu$, we performed an
statistical analysis of the data in the range $8\le W \le 10.5$ with the
Levenberg-Marquardt method for nonlinear least-squares models.
The most likely fit is determined by minimizing the $\chi^2$ statistic
of the fitting function, which we choose to be of the form
\begin{equation}
{\xi_M\over M}=\sum_{i=0}^3 A_i(W-W_{\rm c})^iM^{i/\nu}.
\end{equation}
The critical disorder found for $u=1$ is $W_{\rm c}=9.3\pm
0.2$, and the corresponding critical exponent is equal to $\nu=2.4\pm 0.5$.
The error bar results mainly from the uncertainty in the critical
disorder.

The amount by which we must shift the raw data of Fig.\ 2 to get the 
universal curve of Fig.\ 3 gives us the scaling parameter $\xi$ as a
function of disorder for TIP.
In the inset of Fig.\ 3 we plot the disorder dependence of $\log\xi$ for 
$u=1$ and the center of the band ($E=0$). 

It should be emphasized that we only demonstrate the existence of
correlated two-particle extended states. Probably, there exist many more
uncorrelated localized states where the two particles are away from each
other. Our procedure picks up the longest decay length, at the energy
considered, which is associated with the most delocalized states.

\begin{figure}
\begin{picture}(220,200) (-65,-50)
\epsfbox{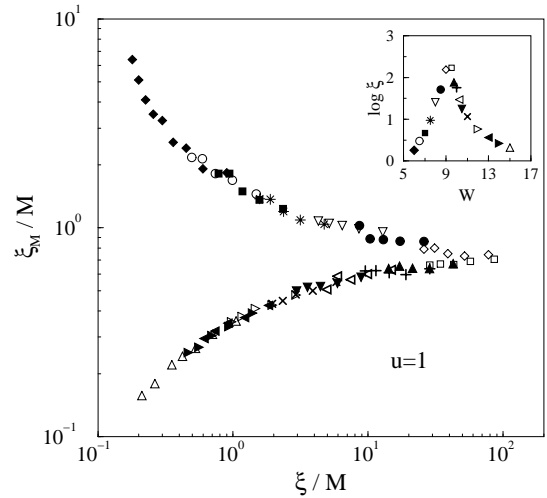}
\end{picture}
\caption{Log-log plot of $\xi_M/M$ as
a function of $\xi/M$ for the data represented in Fig.\ 2. 
Inset: disorder dependence of the scaling parameter $\xi$.
\label{fig.3}}
\end{figure}

The results for $u=0$ are qualitatively different from those for $u=1$.
In the non-interacting case, $\xi_M/M$ decreases with increasing $M$ for
all values of the disorder $W$ considered, and so there appears only one 
branch in the scaling procedure. In Fig.\ 4 we represent  
$\xi_M/M$ as a function of the scaling parameter $\xi$ divided
by $M$. The upper inset shows the disorder dependence of  $\log\xi$.
For comparison we also represent the one-particle 
localization length divided by two $\xi_1/2$ (solid line), which 
is the expected result for relatively strongly localized systems at $u=0$.
We take for $\xi_1$ the value reported by MacKinnon and Kramer \cite{MK83}.
The agreement between our results and $\xi_1/2$ is a positive check of 
the validity of our method of calculation.
In the range of validity of our results we do not obtain any artificial
transition for non-interacting particles, as previously reported for one
particle in 2D \cite{PS81}, due to a different interpretation of the raw 
data. We use MacKinnon and Kramer's \cite{MK83} 
interpretation, which produces no artificial transition either for the
one-particle problem or for our two-particle states calculations.
As a further check, we have applied our method to the well studied 1D
problem. In the lower inset of Fig.\ 4 we show $\xi$ versus disorder on
a double logarithmic scale for two interacting (solid circles) and two
non-interacting (empty circles) bosons in 1D. The straight line
corresponds to $\xi_1/2$, where the one-particle localization length
$\xi_1$ is taken equal to $105/W^2$. We consider samples with 500 sites.
Our results agree with well established previous calculations 
\cite{JS95,SK97,LR98}.

The extension of our results to the case of degenerate electrons, so that
can be applied to explain the transitions found experimentally in Refs.\
\cite{KS96} and \cite{PF97} and to the scaling theory of localization
including interactions \cite{Fi83}, is a very difficult problem.
In three-dimensional systems, Imry \cite{Im95} argued about the 
existence of an effective two-particle mobility edge that would approach
the Fermi energy faster than the single-particle mobility edge.
Our results in 2D could be interpreted as due to the existence of a 
two-particle mobility edge which overcomes the single-particle mobility 
edge at $W=0$.
Before arguing in favor of a transition in 2D degenerate disordered
systems, one should estimate the lifetime of the coherent pairs.
It is not clear how fast our two-particle states would decay to lower
energy excitations. In the case of degenerate electrons with long-range 
Coulomb interactions, Talamantes {\it et al.} \cite{TP96} reported an 
increase of the localization length with respect to the non-interacting 
case, but they only considered the strongly localized regime. 

\begin{figure}
\begin{picture}(220,200) (-65,-50)
\epsfbox{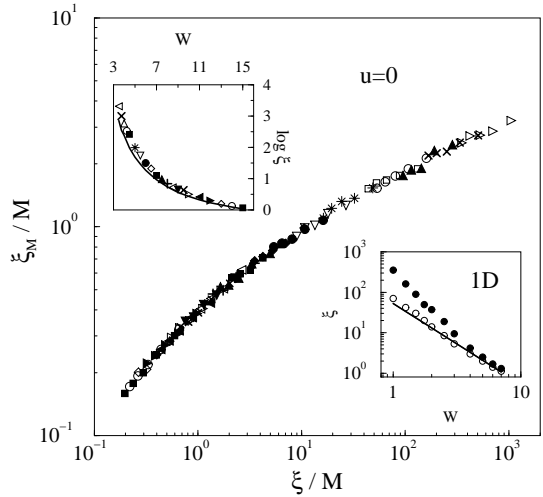}
\end{picture}
\caption{Log-log plot of $\xi_M/M$ as a function of $\xi/M$ for
the non-interacting case. Upper inset: disorder dependence of  
$\log\xi$ along with the one-particle localization length divided 
by two, $\xi_1/2$ (solid line). Lower inset: $\xi$ versus $W$ for 
two interacting (solid circles) and two non-interacting (empty circles) 
bosons in 1D. The straight line corresponds to $\xi_1/2=105/2W^2$.
\label{fig.4}}
\end{figure}

To summarize, we have calculated numerically the decay length of TIP 
in 2D disordered bars of several widths.
The results are consistent with the assumption of a scaling hypothesis. 
Through a scaling analysis of
the data for $u=1$, we proved that there is a localized to extended 
transition at a critical disorder $W_{\rm c}\approx 9.3$.
The critical exponent for the localization length is $\nu\approx 2.4$.
Our method 
clearly indicates the existence of delocalized states for small 
disorders, although the values of the critical disorder and 
exponent can appreciably change when larger system sizes can be 
handled.

We would like to acknowledge financial support from the Spanish DGES, 
project number PB96-1118.

\end{document}